# GRANULAR MEDIA UNDER VIBRATION IN ZERO GRAVITY: TRANSITION FROM RATTLING TO GRANULAR GAS

**P. Evesque[1], Y. Garrabos[2], G. Zhai[3], M. Hou[3]**

[1] Lab MSSMat, UMR 8579 CNRS, Ecole Centrale Paris, 92295 Châtenay-Malabry, France, e-mail: pierre.evesque@ecp.fr

[2] ESEME, ICMCB-cnrs, 33608 Pessac cedex, France ; garrabos@icmcb.u-bordeaux.fr

[3] Institute of Physics, Chinese Academy of Sciences, Beijing 100080, China; mayhou@aphy.iphy.ac.cn

**Abstract:**
*We report on different experimental behaviours of granular dissipative matter excited by vibration as studied during the 43rd ESA campaign of Airbus A300-0g from CNES. The effect of g-jitter is quantified through the generation of a rattle effect. The French-European team's electromagnetic set-up is used, with 20Hz cam recording and high speed camera for a short duration (1s) during each parabola.*

**Pacs # : 5.40 ; 45.70 ; 62.20 ; 83.70.Fn**

## Introduction:

The experimental goal was to address the physics of dissipative systems with excitation of mechanical vibration in zero gravity. We intend to provide answer to questions such as: Is the physics of a collection of grains extensive? How the mechanical excitation is communicated to the grains? How does it propagate into the cell? How does it generate segregation of particles or clustering? Two kinds of cells were used, one for preparing a Chinese satellite experiment with a set of 4 quasi-2d cells, the other one to achieve testing the Maxus 7 rocket of the French-European team [1].

## Test of Chinese satellite cell

Some preliminary conclusion about the satellite cell of this campaign is illustrated in the Figures below; it includes: for dense cells particles were expending and occupying the "maximum" available space instead of forming clusters. Using spatial Fourier transform, it was found the persistence of a global crystalline structure in the densest cell (80% filling ratio) in 0g, while this order disappears in 0g in the 60% cell, proving the "liquid- or gas-like" structure. And for the segregation cells we observed large balls forming cluster in rather low concentration of small particles; the cluster position is found to fluctuate with time, probably sensitive to g-jitter.

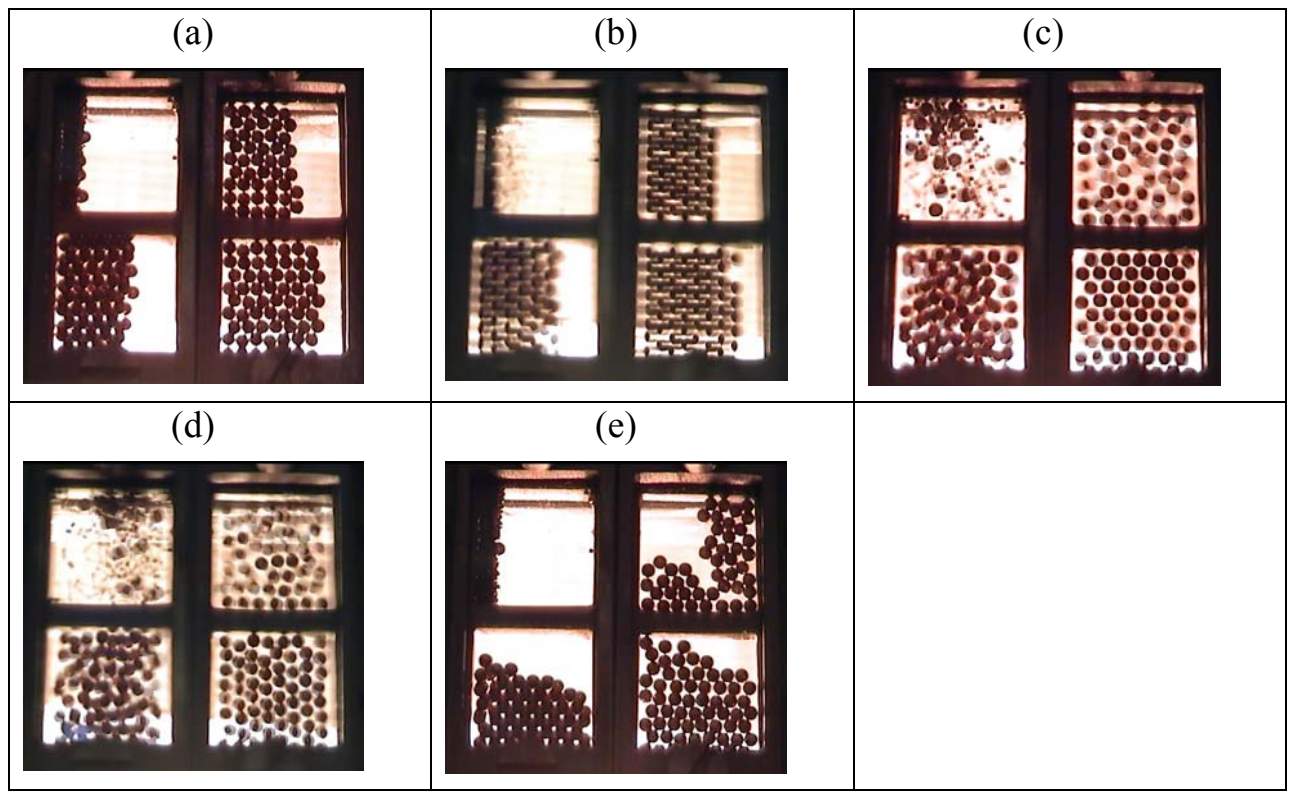

*Figure 1: Chinese satellite cell: (a)* **2g** & **no** vibration*;*
  *(b)* **2g** & vibration: a=0.9 mm, 50Hz, aω²=90m/s²;
  (c) **0g** & vibration: a=0.3 mm, 50Hz, aω²=30m/s²;
  (d) **0g** & vibration: a=0.9 mm, 50Hz, aω²=90m/s²;
  (e) **0g** & vibration: a=6 μm, 130Hz, aω²=4m/s².

Besides, we obtained the acceleration threshold needed to generate the ball vibration in preparation for the capsule experiments. More careful analyses of the data are still undergoing for further understanding of our results.

## Study of g-jitter with the 2d cell of the French-European team (29mm*10mm)

We have tried to study the effect of g-jitter using the 2d sub-cell of the Maxus rocket cell of the French-European team[1]. It has been found that the gas cannot be formed if g-jitter is too large and we see some erratic rattle effect.

On the contrary, when $v_b=a\omega$ is large enough, where a is the amplitude and $f=\omega/(2\pi)$ is the frequency of vibration, gas forms and the g-jitter becomes negligible.

A way to analyse this effect is to compute the position R=(X,Z) of the centre of mass of the gas as a function of time and to measure its standard deviation ΔR. If it is of the order ΔR=L/2 (L being the cell size), we see a rattle effect: the mass centre of the balls goes from one end of the cell to the opposite one. On the other hand, if ΔR is much smaller than L then we get a "gas", whose internal "pressure" stabilises the mass centre in the middle of the cell; and the larger the "pressure" the smaller the fluctuations of ΔR. However ΔR cannot be smaller than the typical fluctuations of the centre of mass of N balls, i.e. $L/(2\sqrt{N})$, (if one excepts cases when some electrostatics takes part). The analysis is done with the cam video.

---

[1] cf Figure 4 after the text, which is added to the original text for Sino-German seminar

In Figure 2 we present different curves representing the Z coordinate of the centre of mass as a function of time during the period of micro-gravity. We see for one of it the rattle effect: Z goes from one edge to the other, for the two others Z remains in the vicinity of the cell centre, which corresponds to a gas-like behaviour.

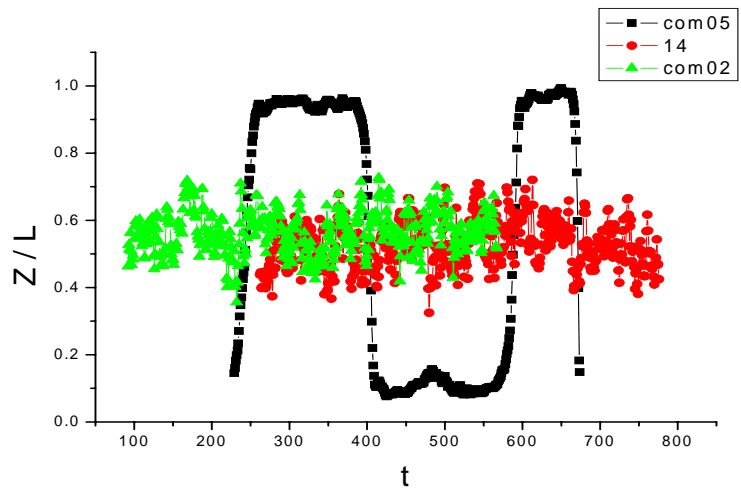

**Figure 2:** position Z of the centre of mass of the "gas" *vs.* time t in frames (1s= 25frames)

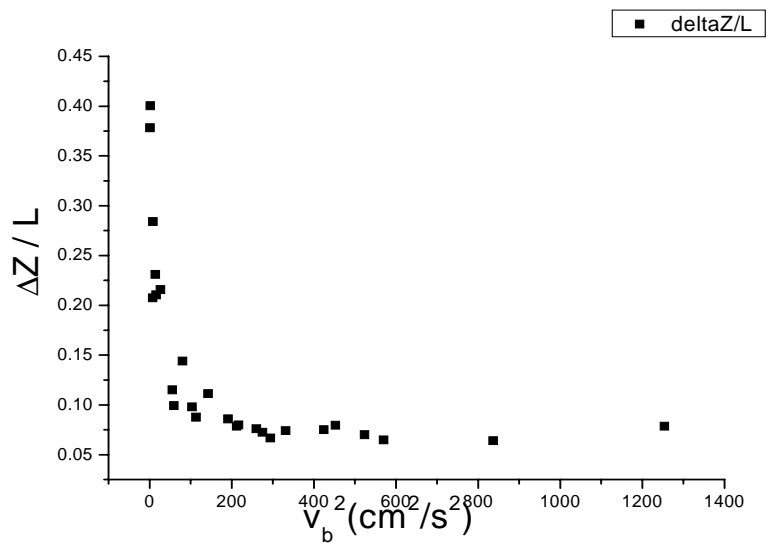

**Figure 3:** Fluctuation ΔZ/L of Z as a function of the driving energy $v_b^2$, ΔZ/L vs. $v_b=a\omega$

In Fig. 3 ΔR is plotted as a function of $v_b^2$. The transition from rattling to granular gas is observed; however, although it is broadened since the real amplitude of g-jitter fluctuates depending on the exact condition of parabola.

*Acknowledgements:* F. Palencia, Z. Sun, H. Yang, T. Zhang, and Novespace are thanked for their assistance. This paper was presented at the 3rd Germany-China Workshop on Microgravity and Space Life Sciences on October 8-11, 2006, at Institute of Physiology, Free University (FU) of Berlin.

## Addendum:

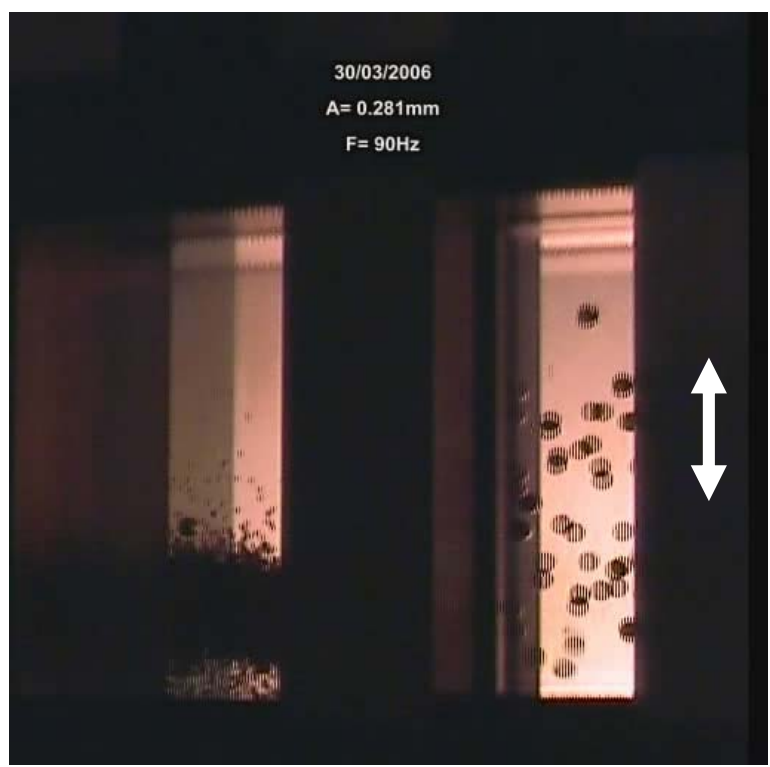

**Figure 4: The Maxus rocket cell recorded via the camcorder during 0g period** in Airbus A300-0g (30/3/2006), amplitude A=0.281mm, frequency f=90Hz ; the 2d cell is the right one. As the camcorder record each frame by recording 2 halve of frames separated by the half duration existing between two frames, each photo appears as the sum on the two complementary half frames. So the balls appear twice separated by the distance it moves during half a frame; the position of the ball is then the average between these two positions. This Fig.4 is added from the original text of the Sino-German seminar.